\documentclass[reprint, superscriptaddress, longbibliography]{revtex4-2}
\pdfoutput=1

\usepackage[dvipsnames]{xcolor}

\usepackage[factor=1100]{microtype}     

\usepackage[font=sf, objectset=centering]{floatrow}
\usepackage{graphicx}
	\graphicspath{{./figures/}}

\usepackage{mathtools}                  
\usepackage{amssymb}                    
\usepackage{amsmath}                    
	
\usepackage{nicefrac}                   
\usepackage[
	mode=text,
	parse-numbers=false,
	parse-units=false,
]{siunitx}                              
\usepackage{braket}                     
\usepackage{diffcoeff}                  

\begin{document}

\title{Additive Laser Excitation of Giant Nonlinear Surface Acoustic Wave Pulses}

\author{Jude Deschamps}
\affiliation{Department of Chemistry, Massachusetts Institute of Technology, Cambridge, MA 02139, USA}
\author{Yun Kai}
\affiliation{Department of Chemistry, Massachusetts Institute of Technology, Cambridge, MA 02139, USA}
\author{Jet Lem}
\affiliation{Department of Chemistry, Massachusetts Institute of Technology, Cambridge, MA 02139, USA}
\affiliation{Institute for Soldier Nanotechnologies, Massachusetts Institute of Technology, Cambridge, MA 02139, USA}
\author{Ievgeniia Chaban}
\affiliation{Department of Chemistry, Massachusetts Institute of Technology, Cambridge, MA 02139, USA}
\author{Alexey Lomonosov}
\affiliation{B+W Department, Offenburg University of Applied Sciences, 77652 Offenburg, Germany}
\author{Abdelmadjid Anane}
\affiliation{Unité Mixte de Physique CNRS/Thales, UMR CNRS 137, 91767 Palaiseau, France}
\author{Steven E. Kooi}
\affiliation{Institute for Soldier Nanotechnologies, Massachusetts Institute of Technology, Cambridge, MA 02139, USA}
\author{Keith A. Nelson}
\affiliation{Department of Chemistry, Massachusetts Institute of Technology, Cambridge, MA 02139, USA}
\affiliation{Institute for Soldier Nanotechnologies, Massachusetts Institute of Technology, Cambridge, MA 02139, USA}
\author{Thomas Pezeril}
\email[Corresponding author\\]{pezeril@mit.edu}
\affiliation{Department of Chemistry, Massachusetts Institute of Technology, Cambridge, MA 02139, USA}
\affiliation{Institut de Physique de Rennes, UMR CNRS 6251, Université Rennes 1, 35042 Rennes, France}

\begin{abstract}

The laser ultrasonics technique perfectly fits the needs for non-contact, non-invasive, non-destructive mechanical probing of samples of mm to nm sizes. This technique is however limited to the excitation of low-amplitude strains, below the threshold for optical damage of the sample. In the context of strain engineering of materials, alternative optical techniques enabling the excitation of high amplitude strains in a non-destructive optical regime are seeking. We introduce here a non-destructive method for laser-shock wave generation based on additive superposition of multiple laser-excited strain waves. This technique enables strain generation up to mechanical failure of a sample at pump laser fluences below optical ablation or melting thresholds. We demonstrate the ability to generate nonlinear surface acoustic waves (SAWs) in Nb:SrTiO$_3$ substrates, at typically \SI{1}{kHz} repetition rate, with associated strains in the percent range and pressures close to 100~kbars. This study paves the way for the investigation of a host of high-strength SAW-induced phenomena, including phase transitions in conventional and quantum materials, plasticity and a myriad of material failure modes, chemistry and other effects in bulk samples, thin layers, or two-dimensional materials.

\end{abstract}

\maketitle


Surface acoustic waves involving nanometer-scale atomic motions are used with routine, low-power operation in a plethora of microelectronic devices~\cite{SAW_review}. The confinement of strain near the sample surface and propagation over millimeter distances make SAWs attractive for a wide range of applications. Typically, these MHz-frequency waves are coherently generated and detected by interdigital microwave transducers (IDTs) deposited on a piezoelectric layer or substrate. Though well-established IDT SAW techniques are limited by small strain amplitudes, it is already possible to employ them to tune material properties~\cite{SAW_review, SAW_spin_review, Foerster_2017, Iikawa_2019, Fandan_2020}. Reaching much higher SAW strain amplitudes, up to the mechanical failure of a sample, would open a wide range of novel applications. Until now, pulsed lasers have been used to generate high-amplitude nonlinear SAW shocks with strain amplitudes in the percent range and pressures in the 10~kbar range \cite{Lomonosov_1999, Lomonosov_2002, Lomonosov_2004, Lomonosov_2008}, but only operating on a single-shot basis because optical damage to the irradiated sample region which occurs even if the shock itself does not cause damage where it propagates.

Large static strains of several percent can be applied to bulk or thin samples up to the onset of plasticity or fracture~\cite{Xu_2020, Hameed_2021, Schlom_2014, Yu_2014, Dhole_2022}. However, since kinetic effects can lead to distinct differences from quasi-static behavior, approaches capable of generating fast actuation mechanisms to explore material responses to dynamic strain are of keen interest. For that purpose, laser-based methods for generation of shock waves have been developed for the study of samples under intense dynamic strain loading. In the most common configuration, bulk shock waves are studied, sometimes non-destructively from femtosecond lasers with a maximum pressure in the 10~kbar range \cite{Capel_2006, Bojahr_2012, Temnov_2013, Pezeril_2014, Pezeril_2015}, but most often destructively in the nanosecond regime with pressures reaching hundreds of
~kbars \cite{Pezeril_2011, Morard_2018, Brenna_2019, McBride_2019, Dlott}. Optical damage imposes severe restrictions including a very limited number of shots on a sample of small size or in a specialized environment such as a cryostat, poor shot-to-shot reproducibility for samples that are not uniform from one region to the next, and no possibility for many measurements on the same sample region in order to explore cumulative effects of repeated moderate shock loading.

Here, we present a methodology for the excitation of non-destructive nonlinear SAWs in Nb:SrTiO$_3$ substrates, at high-repetition rate, limited only by the mechanical strength of the sample. Our technique for the excitation of nonlinear strains in the percent range is based on the spatio-temporal superposition of numerous laser-excited nanosecond strain waves for SAW amplification. In the fundamental point of view, we demonstrate herein a different way to reach nonlinear acoustics. Instead of using a single high-amplitude strain wave already in the nonlinear regime, we show that linear superposition of many weak strain waves can build up a shock wave in a solid. Additionally, we present a femtosecond defocusing imaging technique that can adequately reveal with great sensitivity the nonlinear reshaping of the propagative SAW.


\begin{figure*}[!htp]
	\includegraphics[page=1, width=0.85\columnwidth]{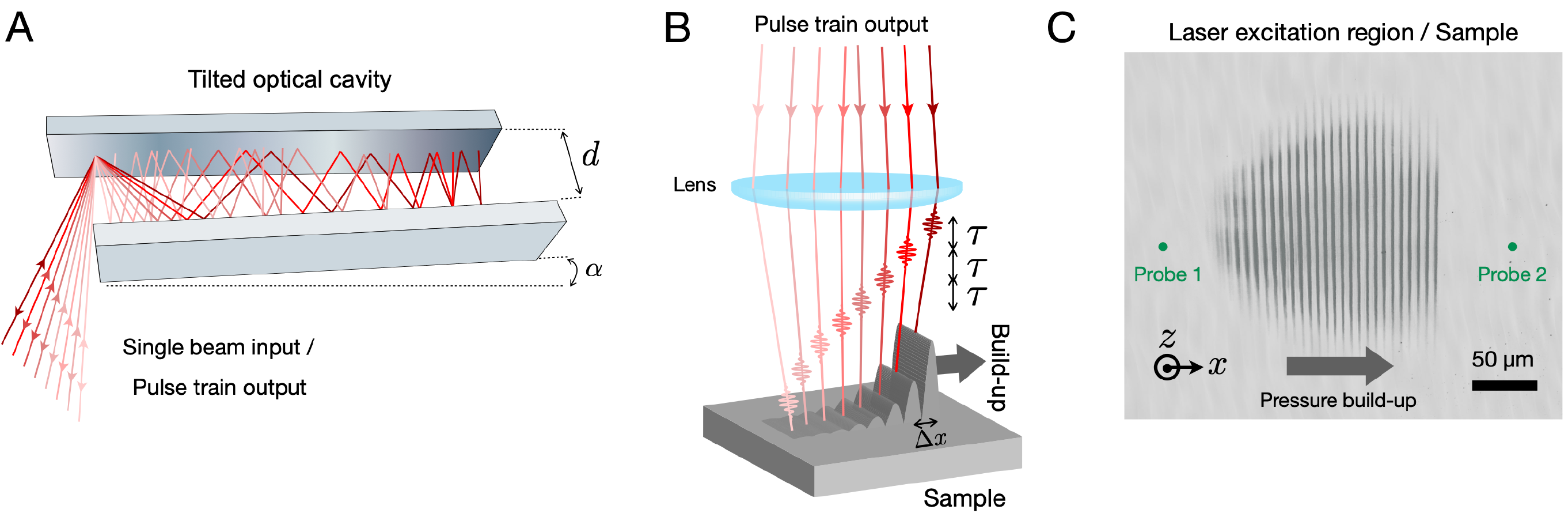}
	\caption{Schematic representation of the platform. (A)~A pulsed laser beam, focused by a cylindrical lens into a tilted optical cavity, generates a pulse train with a temporal separation~$\tau=2d/c$, in the nanosecond range. (B)~The pulse train is focused on the sample surface; each laser pulse is laterally displaced by an amount~$\Delta{x}$ to match the propagation of the generated acoustic wave. The lateral displacement is fine tuned by adjusting the mirror misalignment angle~$\alpha$. (C)~Representative image of the sample surface, showing the laser excitation profile of the pulse train and the location of the two interferometric probes.}
	\label{Fig1}
\end{figure*}

Various configurations enabling the spatio-temporal tailoring of laser sources for acoustic wave generation in the MHz frequency range have been explored in the past, ranging from a moving laser source~\cite{Berthelot_1987}, to an array of synchronized lasers~\cite{Fink_1993, Murray_1996}, to the use of a white cell to split an input laser beam~\cite{Deaton_1994}. The current approach borrows the general idea while simplifying the required experimental apparatus. The set-up used in this study employs a free-space angular-chirp-enhanced delay (FACED) cavity~\cite{Wu_2016}, depicted in Fig.~\ref{Fig1}A, whose purpose is to split an input pulsed laser beam (300~picoseconds pulse duration, \SI{1}{kHz} repetition rate) into a train of sub-pulses spread in time and space. The device consists of two slightly angled high-reflective planar mirrors separated by a distance~$d$  corresponding to an inter-pulse time separation~$\tau \approx 2d/c$, where~$c$ is the speed of light. In this work, $d$ is about 30~cm, corresponding to an inter-pulse time separation $\tau$ of 2 nanoseconds. Since the train of pulses exiting the FACED device are angularly dispersed, each output beam of the pulse train can be focused to a separate location on the sample surface, as depicted schematically in Fig.~\ref{Fig1}B. An array of line-sources is obtained in the end at the sample surface, as shown on the image of~Fig.~\ref{Fig1}C. See SI Appendix for the full optical schematic. 

Each individual line gets absorbed in the opaque material, inducing a thermoelastic response. This process launches acoustic waves that propagate in the vicinity of the surface, namely a surface acoustic wave~(SAW) and a surface-skimming longitudinal wave~(SSLW), moving at the Rayleigh wave velocity~$v_R$ and longitudinal wave velocity~$v_L$, respectively. By tuning the tilt angle~$\alpha$ of the FACED cavity, the line spacing $\Delta{x}$ can be finely scanned until the spatio-temporal spread matches the propagation velocity of either wave. Note that the tilt angles are so small ($\sim$mrad) that tuning them does not substantially modify the inter-pulse delay~$\tau$. At specific angles~$\alpha$, the coherent superposition of all the individual acoustic waves leads to the build-up of a large-amplitude wave in the phase-matched direction. The surface displacement~$u_z$ induced by the passage of the acoustic waves is monitored through reflective interferometric probing~\cite{Glorieux_2004} using two optical probes located on both sides of the excitation region,  see Fig.~\ref{Fig1}C.

\begin{figure}[!h]
	\includegraphics[page=1, width=0.9\columnwidth]{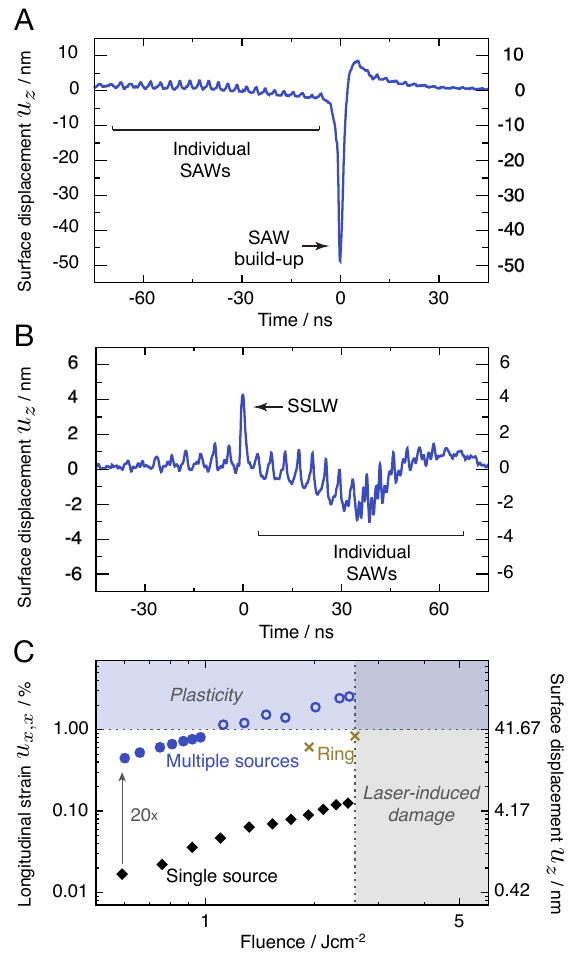}
	\caption{Time traces of the interferometric displacements $u_z$ and corresponding calculated in-plane strains $u_{x,x}$. (A) Amplification of the SAW for a Nb:STO sample, taken at~\SI{1.0}{J~cm^{-2}} laser fluence. (B)~Same as~(A) but for a SSLW. (C)~Comparison of the current multi-line approach (blue circles) to a single-line generation scheme (black diamonds). On Nb:STO, the multi-line approach generates SAWs that can bring the material into the plastic regime (empty circles), with mechanical failure occurring from cumulative fatigue. The range of surface displacements and corresponding strains achievable can be extended by lowering the repetition rate of the SAW generation, from the repetitive 1~kHz regime (full circles) to the finite-shot regime (empty circles). For Nb:STO, a comparison to the ring shock geometry \cite{Pezeril_2011} is also shown (crosses). The colored regions denote the regimes where mechanical and/or optical damage are observed on the sample during or after laser irradiation.}
	\label{Fig2}
\end{figure}

In order to demonstrate the effectiveness of the acoustic amplification platform, we performed experiments on Nb doped SrTiO$_3$ substrates (Nb:STO). Above all, SrTiO$_3$ is among the most common substrate for sample deposition, and as such could be used in the study of strain-induced effects in materials that are either deposited as exfoliated 2D layers or grown as thin films on the substrates. Figure~\ref{Fig2}A and B shows typical waveforms acquired from the Nb:STO sample for parameters tuned to amplify either the SAW moving at the Rayleigh wave velocity $v_R$ or the surface-skimming longitudinal wave (SSLW) moving at the longitudinal wave velocity $v_L$. Not that the additive amplified wave is detected at the probe~2 location and non-phase-matched waves, travelling in the opposite direction, are recorded by probe~1, see Fig.~\ref{Fig1}C. The huge amplified SAW pulse is very prominent as compared to the tiny ripples that come from all the individual SAWs travelling in the opposite direction. The SSLW is significantly amplified as well, but, since the interferometric detection is sensitive to the surface displacement which is perpendicular to the longitudinal displacement, it is hidden in the SAW ripples.

\section*{Results and Discussions}

\subsection*{Train pulse excitation vs single pulse vs ring excitation}

A comparison of SAW generation in the current multi-line method to a single-line approach is shown in Fig.~\ref{Fig2}C. Very high interferometric displacements $u_z$ close to the 100 nanometers range, and corresponding in-plane strain amplitudes $u_{x,x}$, can be reached using the multi-lines approach with a Nb:STO substrate. As shown on the left vertical scale of Fig.~\ref{Fig2}C, strain amplitudes of~\SI{1.0}{\%} can be attained at a~\SI{1}{kHz} repetition rate, with strains up to~\SI{3.0}{\%} observed in the finite-shot regime\,---\,when the total number of shots is limited to several hundreds. These high amplitudes stem from a careful choice of Nb doping level of ~\SI{0.7}{\%} to make the optical penetration depth match the SAW depth profile of the material. Importantly, multi-line excitation in the finite-shot regime enables the onset of plasticity to be reached before any optical damage is induced from the pump laser light in the irradiated region, which is not the case for a single line excitation with about 20$\times$ lower amplitude. In this plastic regime, corresponding to the fluence range from 1.1 to \SI{2.6}{J~cm^{-2}} and corresponding SAW strains above~\SI{1.0}{\%}, each laser shot induces dislocations and plastic deformations that alter the mechanical integrity of the sample. 

High strain amplitudes can alternatively be achieved through the use of a ring excitation geometry, which enables acoustic amplification through a focusing geometry. For comparison with the multi-line approach, we have performed ring shock experiments on Nb:STO substrates under the same experimental conditions as in~\cite{Pezeril_2011}. For a given fluence, strain amplitudes about three times higher can be attained with the multi-line 1D geometry compared to the 2D single ring geometry, see Fig.~\ref{Fig2}C. In addition to its increased efficacy, the current 1D planar scheme simplifies the interpretation of the data since the strain is unidirectional and there is no acoustic discontinuity from focusing. In addition, the multi-line scheme is better adapted for the incorporation of complimentary techniques, such as leads for conductivity measurements, as the strain pulses are delivered to a region decoupled from the excitation region.

\subsection*{Shock-induced mechanical damage without optical damage}

\begin{figure}[!tbp]
	\includegraphics[page=1, width=0.9\columnwidth]{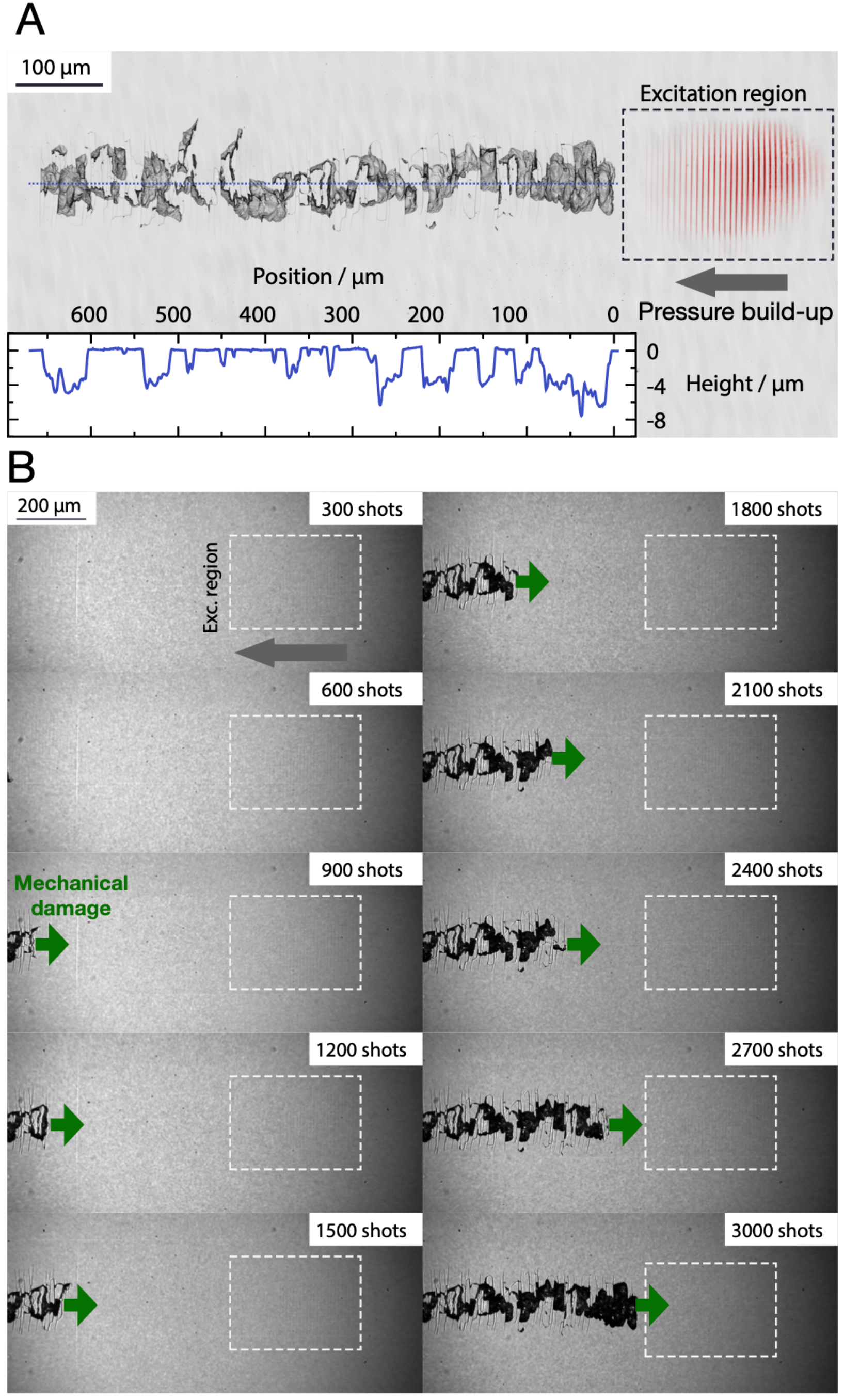}
	\caption{Shock-induced mechanical damage without optical damage. (A)~Confocal image of the surface of a Nb:STO(100) sample showing SAW-induced mechanical damage hundreds of microns away from the laser-excitation region. The SAW propagates along the [010] direction. Inset: depth profile along the dotted line.(B)~Snapshots of SAW-induced damage on Nb:STO(100). It takes about 3,000 shots for the mechanical damage to reach the laser-excitation region. }
	\label{Fig3}
\end{figure}

In the fluence range from 1.1 to \SI{2.6}{J~cm^{-2}} that corresponds to the plastic regime, repetitive laser shots, in the range of several hundreds, lead to mechanical failure from cumulative fatigue. Fig.~\ref{Fig3}A shows a \emph{post-mortem} confocal microscopy image of a Nb:STO surface taken following an experiment in the plastic regime, at a pump fluence of~\SI{1.1}{J~cm^{-2}}. The laser excitation sequence was interrupted after 3,000~shots, before the mechanical damage reached the excitation region. Fracture and dislocations in the substrate, along with regions of material ejection, can be seen a few hundreds of microns away from the non-damaged pump region, in the direction of propagation of the amplified SAWs. 

In order to monitor the evolution of the SAW-induced mechanical damage on a Nb:STO(100) sample, we have used a femtosecond light pulse to get an instantaneous snapshot image of the sample surface following laser-excitation. The laser fluence of the train pulse was set to~\SI{1.1}{J~cm^{-2}} and the laser repetition rate was lowered to~\SI{100}{Hz} for these experiments. As shown in Fig.~\ref{Fig3}, snapshot images obtained at intervals of 300 pump shots reveal that after about 600~shots, the damaged region becomes visible within the camera's field of view, and it eventually reaches the excitation region after 3000 shots. This observed trend reveals that mechanical damage initially occurs at the location farthest from the excitation region, with subsequent shots progressively inducing fractures closer to the excitation region. 

It is worth emphasizing that a substantial number of shots, in the thousands range, is necessary to induce mechanical alterations on the sample surface in the immediate vicinity of the laser excitation region. This particular region holds significant importance as it allows for the deposition and testing of ultra-thin materials under high strains, without being subjected to light pulses. The requirement of such a significant number of shots, exceeding what is typically achievable in a conventional single-shot laser-shock experiment, represents a highly promising result within the field of strain engineering.

\subsection*{Non-linear SAW observation at 1~kHz and modeling}

\begin{figure*}[!tp]
\begin{center}
    	\includegraphics[page=1, width=0.9\columnwidth]{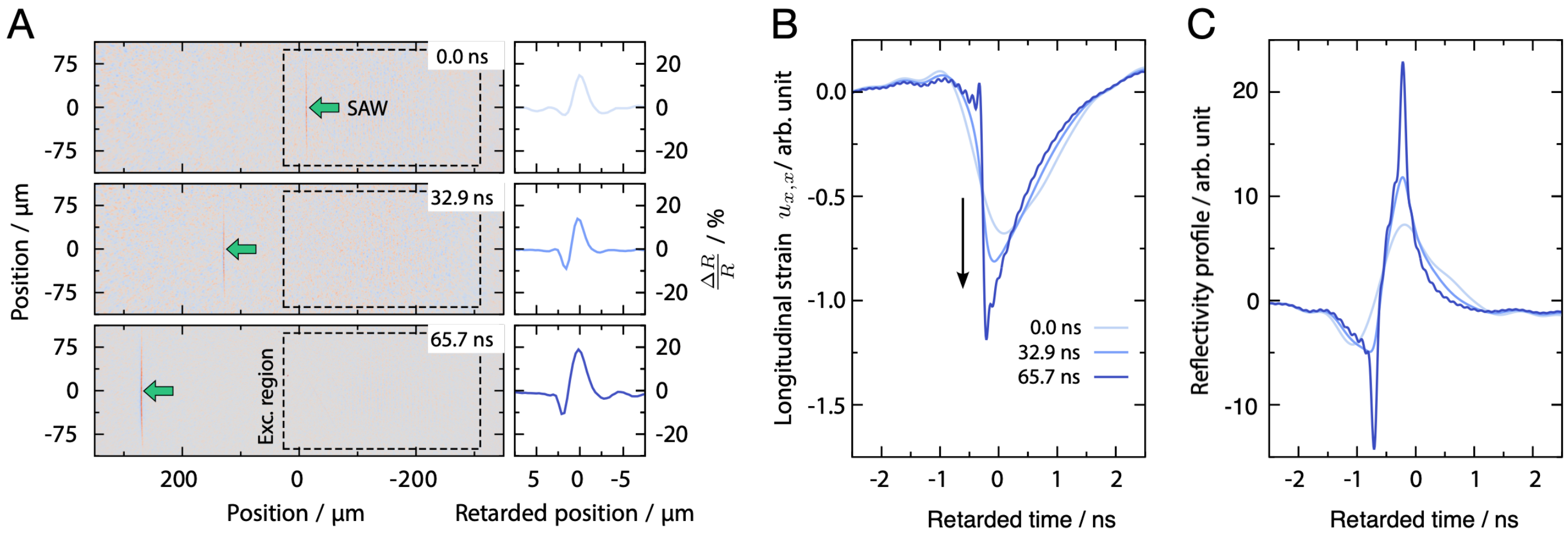}
     \end{center}
	\caption{Non-linear reshaping of the SAW. (A) Time-resolved femtosecond reflectivity images for SAWs traveling along the~$\text{[}\overline{\text{11}}\text{2]}$ direction of a Nb:STO(111) sample. An arrow points to the SAW location in each frame. The excitation region is outlined with a dashed box. Profiles extracted from a cut of the reflectivity maps along the horizontal dimension are shown on the right. (B) Numerical modeling of the longitudinal strain $u_{x,x}$ that matches the experimental observation in (A). The arrow shows the increase in peak strain as the wave propagates. (C)~Corresponding simulated optical reflectivity profiles.}
	\label{Fig4}
\end{figure*}

The onset of plasticity and mechanical failure from the multi-line scheme can be explained by analyzing the nonlinear evolution of the SAWs as they propagate. High-amplitude waves will develop a steepening shock front with increasing peak stress as they propagate due to the elastic nonlinearity of the medium~\cite{Kalyanasundaram_1982, Parker_1988, Lomonosov_1999, Hamilton_1999}. The nonlinear SAW reshaping can be revealed by techniques sensitive to a spatial derivative of the waveform, such as Schlieren imaging or the shadowgraph technique~\cite{Settles_2001}. In this work, a simple transient reflectivity set-up with a slightly defocused imaging system has been used to probe the SAW temporal evolution. Fig.~\ref{Fig4}A shows a series of snapshots of the  Nb:STO(111) sample surface, using a femtosecond light pulse probe as the illumination source. The direction of propagation was chosen to be~$\text{[}\overline{\text{11}}\text{2]}$ to maximize the elastic nonlinearity~\cite{Lomonosov_2004, Hess_2005}. The contrast~$\Delta{R}/R$ in the transient reflectivity maps of Fig.~\ref{Fig4}A stems from two contributions, namely the photoelastic effect~\cite{Yamazaki_2004} and Fresnel diffraction from propagation of the optical field over the defocus length~$\Delta{z}$. The latter effect has been analyzed before in the context of phonon-polariton imaging~\cite{Koehl_1999}. For small propagation lengths of the optical field, it is most sensitive to the Laplacian of the surface displacement (see SI Appendix):
\begin{equation}
	\frac{\Delta{R}}{R}
	\propto - \frac{\lambda \Delta{z}}{2\pi} \nabla^2{u_z(x,y)},
\end{equation}
with~$\lambda$ the probe wavelength. The contribution of defocusing to the measured contrast can reach tens of percent, dominating over the photoelastic contribution, easily revealing the location and shape of the acoustic wave. The predominant effect of the Fresnel diffraction has been further confirmed experimentally from the fact that the reflectivity contrast switches sign when $\Delta{z}$ is reversed.

From the evolution of the reflectivity data taken as the SAW moves away from the excitation region, shown in Fig.~\ref{Fig4}A, we experimentally evidence the reshaping of the SAW profile. The data show a significant growth of the positive reflectivity peak and of the negative peak, initially absent at 0.0~ns (this time is taken as the time the amplified SAW leaves the excitation region). This nonlinear evolution is confirmed by numerical simulations obtained from propagating an experimental interferometric displacement waveform at delays matching those of Fig.~\ref{Fig4}A (see SI Appendix). The remarkable novelty here is the experimental observation of the nonlinear reshaping of SAWs in a repetitive and non-destructive manner. It demonstrates the possibility to average repetitive multi-shot experiments in the non-linear regime. Previously, the observation of Rayleigh shock waves had been limited to single-shot experiments \cite{Lomonosov_1999, Lomonosov_2002, Lomonosov_2004, Lomonosov_2008}. The implementation of a high kHz repetition rate in this research holds great promise for significantly enhancing the signal-to-noise ratio of the collected data, which represents a substantial leap forward in the field.

Fig.~\ref{Fig4}B shows the evolution of the longitudinal strain $u_{x,x}$ at the free surface, extracted numerically from the particle displacements measured experimentally in Fig.\ref{Fig2}A. These simulations allow to draw some conclusions on the non-linear reshaping of the SAW. Initially, as expected from the thermoelastic excitation process ~\cite{Scruby}, the strain profile is purely compressive (there is no tensile strain component). Then, as the SAW propagates, the strain profile remains unipolar with a compressive strain component that significantly grows in amplitude. Note that in terms of strain-engineering applications, it can be convenient to load samples with high-amplitude unipolar strains and not a mix of positive and negative strains that are more complex to disentangle. The projection of the stress along particular slip systems eventually reaches the yield stress of the material, which explains how fracture first appears significantly away from the excitation region when the pump fluence is set just above the onset of plasticity. Fig.~\ref{Fig4}C shows the second derivative of the surface displacement, qualitatively matching the evolution of the reflectivity profiles shown in Fig.~\ref{Fig4}A, which further confirms that nonlinear dynamics are captured from time-resolved reflectivity measurements. Note that the simulation in Fig.~\ref{Fig4}C shows a sharper shock front than the results in Fig.~\ref{Fig4}A because of the limited spatial resolution of our optical measurements. 

\section*{Summary and Outlook}

To summarize, this work establishes a versatile non-destructive technique for the generation of high-amplitude SAW strains in solid substrates. Notably, Nb:STO substrates, commonly used for sample deposition, make it possible to reach SAW strain amplitudes of more than~\SI{1.0}{\%} (up to \SI{3.0}{\%} or \SI{100}{kbar} in the finite-shot regime) before any type of damage\,---\,mechanical or laser-induced\,---\,occurs. The current technique enables transduction of high amplitude SAWs into samples of interests deposited on the substrates, and will thereby facilitate the study of dynamic strain effects in correlated materials and other types of samples sensitive to strains. Additionally, the large surface displacements reached using the current platform enable the use of a simple transient reflectivity defocusing imaging technique to track linear or nonlinear acoustic waves in materials with great sensitivity. Extension of this work to bulk compressional waves will be possible with less strongly absorbing samples including STO with a lower Nb doping concentration for reduced absorption of the  pump light. More uniform light absorption through tens of microns (\emph{e.g.} optical penetration depth of roughly 100~$\mu$m will produce a thermoelastic stress that will drive a primarily compressional shock, as we have seen for shock generation with single intense pulses in the single-shot regime~\cite{Pezeril_2011}. Despite their well-known influence on the order parameter in solid state physics and their coupling to correlated electron systems, shear acoustic waves are generally very challenging to study\,---\,primarily very difficult to efficiently laser-excite~\cite{PEZERIL2016}. The excitation of Stoneley or Scholte surface-interface waves as in~\cite{Glorieux_2006}, between Nb:STO and thick solid or viscoelastic liquid materials using the multi-source technique could enable strong shear strain loading and the study of shear degrees of freedoms in disordered or ordered systems (supercooled liquids and glasses, mixed ferroelectrics and multiferroics, magnetic samples).


\subsubsection*{Sample Preparation}
Nb doped STO substrates (\SI{0.7}{\%} wt Nb) were purchased from MTI Corporation. No additional coatings were deposited on the substrates. The high amplitudes SAWs observed in our study, stem from a careful choice of Nb doping level to make the optical penetration depth match the SAW depth profile of the material. A penetration depth of about~\SI{15}{\micro m} for the 800-nm pump light was obtained from a~\SI{0.7}{\%} Nb doping.


\subsubsection*{Optical setup}

\textbf{Pump path}.The pump laser is derived from the uncompressed output of a \SI{1}{kHz} Ti:Sapphire regenerative amplifier (Coherent Astrella) with a central wavelength of \SI{800}{nm} and pulse duration of \SI{300}{ps} FWHM. Up to \SI{1.6}{mJ} of pulse energy is used as input for the tilted optical cavity. A half-wave plate and polarizing beam splitter combination is used as a variable attenuator to tune the pump energy. The pump beam is focused by a cylindrical lens near the back mirror of the cavity, propagates back and forth through the cavity, and reemerges at the entrance location. The output beam from the cavity is reflected by a beam-splitter and is directed towards the sample. Each ray forming the output beam experiences a number of reflections that depends on its incidence angle on the cavity. Consequently, the output of the device is a spatio-temporally spaced pulse train with an inter-pulse separation approximately equal to the round-trip time of the cavity. A second cylindrical lens (orthogonal to the first one) is used to make an array of lines at the focus of the microscope objective that coincides with the sample surface location. See SI Appendix for a detailed sketch of the optical setup. 

\textbf{Probe path}. The probe beam is a continuous-wave frequency-doubled Nd:YAG (Coherent Verdi) laser operating at a central frequency of~\SI{532}{nm} with an average power at the sample lower than~\SI{10}{mW}. The probe first diffracts on a transmissive binary phase mask. The optical field at the phase mask location is then relayed to the back focal plane of the objective lens by a Keplerian telescope; a beam block is placed at the Fourier plane of the telescope to filter out all diffraction orders besides the $\pm 1$~orders. The probes are focused on the sample on each side of the excitation region, as shown in Fig.~\ref{Fig1}C. Their reflections are recombined interferometrically on the same phase mask, and the intensity of the signal beam is detected using a fast avalanche photodiode (Hamamatsu APD C5658) connected to a high-bandwidth oscilloscope. As the amplified surface waves propagate through probe~1, the sample surface displacement~$u_z$ induces an optical path difference~$\xi = 2 u_{z}$ between both arms of the interferometer, which is translated into a change of intensity~$I$ of the signal beam~\cite{Glorieux_2004}:
\begin{equation}
    I(\xi)
    = \Braket{I} \left[1 - V \sin{\left(\frac{2\pi\xi}{\lambda} + \phi_0 \right)}\right],
\end{equation}
where~$\braket{I}$ is the average intensity, $V$~is the interferometric visibility, $\lambda$ is the probe wavelength, and~$\phi_0 \in [-\pi, \pi]$ is the phase offset of the interferometer. Note that the interferometric range is limited to a relative surface displacement~$u_z = \pm\lambda/8 = \pm\SI{66.5}{nm}$ at the highest sensitivity point ($\phi_0 = 0$). However, that range can be extended by tuning the initial phase offset of the interferometer, which is achieved by a horizontal translation of the phase mask.

\textbf{Transient reflectivity imaging path}. Transient reflectivity imaging capabilities are available to the set-up. The compressed output of the Ti:Sapphire regenerative amplifier (Coherent Astrella), with a pulse duration of about \SI{90}{fs} FWHM, is first frequency doubled to \SI{400}{nm} in a barium beta borate (BBO) crystal. A set of variable-length fibers is then used to delay the beam propagation. The beam is finally used as an expanded illumination source on the sample, and its reflection is collected on a CMOS sensor (Hamamatsu Orca Flash) in a stroboscopic fashion. The acquisition of each active frame (with the acoustic wave) is followed by the acquisition of a reference frame (without the acoustic wave).

\textbf{Fluence calibration.} The fluence axis in Fig.\ref{Fig2}C is calibrated using the following procedure. For each data point, the total laser pulse energy that is delivered to the sample is measured. This value is then divided by the total effective area covered by the multi-line or single-line excitation pattern. Note that the effective area of each line is defined as the FWHM of its Gaussian intensity profile.

\subsubsection*{Numerical simulations}

As a short SAW pulse propagates across a nonlinear elastic medium, its shape changes gradually due to the nonlinear distortion. Within the quadratic approximation and assuming the nonlinear distortion is slow (in the sense that the change on a distance comparable with the pulse width is small), the nonlinear evolution can be understood as an interaction among the Fourier components that results in generation of sum and differential frequencies. Quantitatively, this process is described by the evolution equation system derived in~\cite{Eckl_2004}:
\begin{eqnarray}
		i \diff{}{\tau} B_n
		&=& n v_R \, \, [
			\sum_{m = 0}^{n} F(m/n) B_{m}^{\phantom{*}} B_{n-m}^{\phantom{*}} \\ &&+
			2 \sum_{m = n}^{\infty} (n/m) F^*(n/m) B_{m}^{\phantom{*}} B_{m-n}^*
		\, \, ].\nonumber
\end{eqnarray}
Here, $v_R$ is the Rayleigh velocity, $B_n$ is the $n$-th Fourier component of the Rayleigh wave strain profile, $\tau$ is the propagation distance, and the function~$F$ characterizes the rate of harmonics generation.

The experimental waveform registered at the location close to the source is expanded in the Fourier series and fed to the evolution system of equations as the initial condition. The system is integrated numerically over the propagation distance, and finally the inverse Fourier transform gives the predicted waveform. The kernel function~$F$ depends on the second- and third-order elastic constants of the solid, and on the geometry of the problem if the solid is anisotropic. Note that in certain geometries, $F$ may become complex-valued and possess both real and imaginary parts. The second- and third-order elastic constants of the Nb:STO sample used in our numerical simulations were taken from \cite{Lardner_1986, Yang_2013}. See SI Appendix for additional information.

\subsubsection*{Acknowledgement}

We acknowledge insightful discussions with Alexei Maznev from MIT, Andreas Mayer from Offenburg University, Kevin Tsia from the University of Hong Kong, as well as Takakazu Suzuki, visiting student at MIT from Keio University. We acknowledge funding from the DEVCOM Soldier Center and the Assistant Secretary of the Army for Acquisition Logistics and Training, specifically 0601102A Defense Sciences. The MIT experimental work of J.D., Y.K., T.P., and K.A.N. was supported by the U.S. Department of Energy, Office of Basic Energy Sciences, under Award No. DE-SC0019126. T.P. acknowledges financial support from DGA (Direction Générale de l'Armement) under grant ERE 2017 60 0040 as well as from Région Bretagne under grant SAD. Y.K. acknowledges support from a DAAD (German Academic Exchange Service) fellowship.

\bibliography{ms}

\end{document}


\title{Supplemental Material: Additive Laser Excitation of Giant Nonlinear Surface Acoustic Wave Pulses}

\author{Jude Deschamps}
\affiliation{Department of Chemistry, Massachusetts Institute of Technology, Cambridge, MA 02139, USA}
\author{Yun Kai}
\affiliation{Department of Chemistry, Massachusetts Institute of Technology, Cambridge, MA 02139, USA}
\author{Jet Lem}
\affiliation{Department of Chemistry, Massachusetts Institute of Technology, Cambridge, MA 02139, USA}
\affiliation{Institute for Soldier Nanotechnologies, Massachusetts Institute of Technology, Cambridge, MA 02139, USA}
\author{Ievgeniia Chaban}
\affiliation{Department of Chemistry, Massachusetts Institute of Technology, Cambridge, MA 02139, USA}
\author{Alexey Lomonosov}
\affiliation{Scientific and Technological Center of Unique Instrumentation, Russian Academy of Sciences, 117342, Butlerova Str. 15, Moscow, Russian Federation}
\author{Abdelmadjid Anane}
\affiliation{Unité Mixte de Physique CNRS/Thales, UMR CNRS 137, 91767 Palaiseau, France}
\author{Steven E. Kooi}
\affiliation{Institute for Soldier Nanotechnologies, Massachusetts Institute of Technology, Cambridge, MA 02139, USA}
\author{Keith A. Nelson}
\affiliation{Department of Chemistry, Massachusetts Institute of Technology, Cambridge, MA 02139, USA}
\affiliation{Institute for Soldier Nanotechnologies, Massachusetts Institute of Technology, Cambridge, MA 02139, USA}
\author{Thomas Pezeril}
\email[Corresponding author\\]{pezeril@mit.edu}
\affiliation{Department of Chemistry, Massachusetts Institute of Technology, Cambridge, MA 02139, USA}
\affiliation{Institut de Physique de Rennes, UMR CNRS 6251, Université Rennes 1, 35042 Rennes, France}

\maketitle

\begin{figure}[h!]
\centering
\includegraphics[width=0.8\textwidth]{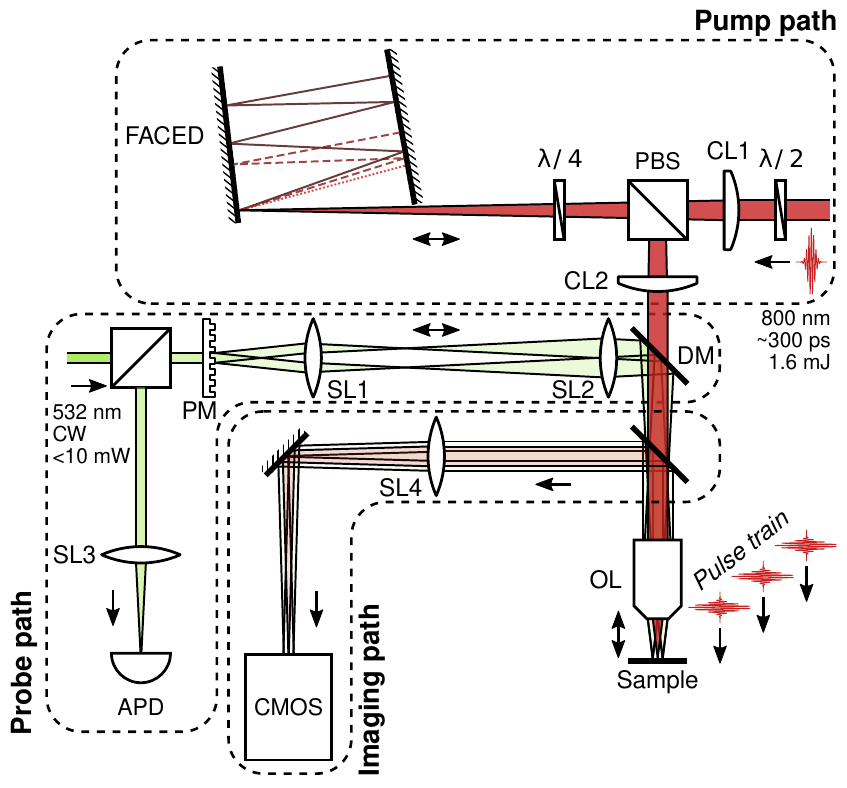}
	\caption{Schematic illustration of the platform. $\lambda/2$: half-wave plate; CL1: cylindrical lens 1 ($f = \SI{500}{mm}$); PBS: polarizing beam splitter; $\lambda/4$: quarter-wave plate; FACED: free-space angular-chirp-enhanced delay device; CL2: cylindrical lens 2 ($f = \SI{500}{mm}$); DM: dichroic mirror; OL: objective lens (Mitutoyo M Plan Apo 10x); PM: binary phase mask; SL1: spherical lens 1 ($f = \SI{100}{mm}$); SL2: spherical lens 2 ($f = \SI{175}{mm}$); SL3: spherical lens 3 ($f = \SI{100}{mm}$); APD: avalanche photo-diode (Hamamatsu C5658); SL4: spherical lens 4 ($f = \SI{400}{mm}$); CMOS: CMOS sensor (Hamamatsu ORCA).}
	\label{fig:schematics}
\end{figure}


\section*{Experimental set-up}
	\label{set-up}
\subsection*{Pump path}

The pump laser is derived from the uncompressed output of a 1~kHz Ti:Sapphire regenerative amplifier (Coherent Astrella) with a central wavelength of \SI{800}{nm} and pulse duration of \SI{300}{ps} FWHM. Up to \SI{1.6}{mJ} of pulse energy is used as input for the FACED cavity. A half-wave plate and polarizing beam splitter combination is used as a variable attenuator to tune the pump energy. {\color{black} The pump beam is focused by the cylindrical lens CL1 near the back mirror of the FACED cavity, propagates back and forth through the cavity, and reemerges at the entrance location. The output beam from the FACED cavity is reflected by the PBS and is directed towards the sample. Each ray forming the output beam experiences a number of reflections that depends on its incidence angle on the cavity. Consequently, the output of the device is a spatio-temporally spaced pulse train with an inter-pulse separation approximately equal to the round-trip time of the cavity, \emph{i.e.}~$\tau \approx 2d/c$, where~$c$ is the light velocity. In this work, $d$ is fixed at about~\SI{30}{cm}, corresponding to an inter-pulse time separation~$\tau$ of~\SI{2}{ns}. A second cylindrical lens CL2 (orthogonal to CL1) is used to make an array of lines at the focus of the microscope objective that coincides with the sample surface location.}

\subsection*{Probe path}
The probe beam is a continuous-wave frequency-doubled Nd:YAG (Coherent Verdi) laser operating at a central frequency of~\SI{532}{nm} with an average power lower than~\SI{10}{mW}. The probe first diffracts on a transmissive binary phase mask. The optical field at the phase mask location is then relayed to the back focal plane of the objective lens by a Keplerian telescope; a beam block is placed at the Fourier plane of the telescope to filter out all diffraction orders besides the $\pm 1$~orders. The probes are focused on the sample on each side of the excitation region. Their reflections are recombined interferometrically on the same phase mask, and the intensity of the signal beam is detected using a fast photodiode connected to a high-bandwidth oscilloscope. As the amplified surface waves propagate through probe~1, the sample surface displacement~$u_z$ induces an optical path difference~$\xi = 2 u_{z}$ between both arms of the interferometer, which is translated into a change of intensity~$I$ of the signal beam~\cite{Glorieux_2004}:
\begin{equation}
    I(\xi)   = \Braket{I} \left[1 - V \sin{\left(\frac{2\pi\xi}{\lambda} + \phi_0 \right)}\right],
\end{equation}
where~$\braket{I}$ is the average intensity, $V$~is the interferometric visibility, $\lambda$ is the probe wavelength, and~$\phi_0 \in [-\pi, \pi]$ is the phase offset of the interferometer. Note that the interferometric range is limited to a relative surface displacement~$u_z = \pm\lambda/8 = \pm\SI{66.5}{nm}$ at the highest sensitivity point ($\phi_0 = 0$). However, that range can be extended by tuning the initial phase offset of the interferometer, which is achieved by a horizontal translation of the phase mask.

\subsection*{Transient reflectivity imaging path}
In addition to the components illustrated in Fig.~\ref{fig:schematics}, transient reflectivity imaging capabilities can be easily added to the set-up. The compressed output of the Ti:Sapphire regenerative amplifier (Coherent Astrella), with a pulse duration of about \SI{90}{fs} FWHM, is first frequency doubled to \SI{400}{nm} in a barium beta borate (BBO) crystal. A set of variable-length fibers is then used to delay the beam propagation. The beam is finally used as an expanded illumination source on the sample, and its reflection is collected on the CMOS sensor in a stroboscopic fashion. The acquisition of each active frame (with the acoustic wave) is followed by the acquisition of a reference frame (without the acoustic wave).

\section*{Acoustic phase matching}

\begin{figure}[!htbp]
\centering
	\includegraphics[width=0.7\columnwidth]{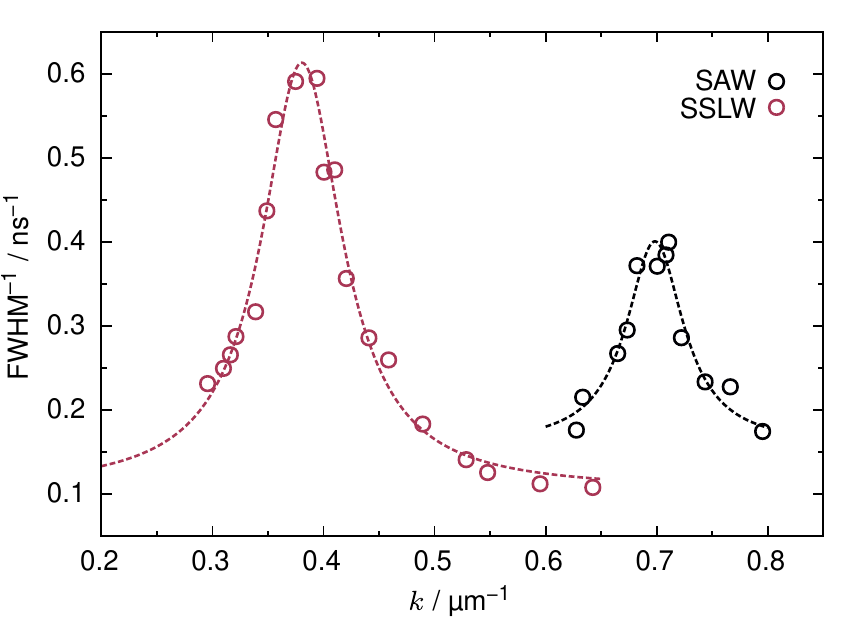}
	\caption{Acoustic resonances obtained on Nb:STO from scanning the line spacing~$\Delta{x}$ of the excitation pattern ($k = 2\pi/\Delta{x}$). Dotted lines are guides to the eye.}
	\label{fig:phase_matching}
\end{figure}

Each individual pump line focused on the sample gets absorbed in the material, inducing a thermoelastic response. This process launches acoustic waves that propagate in the vicinity of the surface, namely a surface acoustic wave~(SAW) and a surface-skimming longitudinal wave~(SSLW), moving at the Rayleigh wave velocity~$v_R$ and longitudinal wave velocity~$v_L$, respectively. By tuning the tilt angle~$\alpha$ of the FACED cavity, the line spacing can be finely scanned until the spatio-temporal spread matches the propagation velocity of either wave. Note that the tilt angles are so small ($\sim$mrad) that tuning them does not substantially modify the inter-pulse delay~$\tau$. At specific angles~$\alpha$, the coherent superposition of individual acoustic waves leads to the build-up of a large-amplitude wave in the phase-matched direction. Fig.~\ref{fig:phase_matching} shows two resonances obtained from scanning the line spacing~$\Delta{x}$ of the excitation pattern, corresponding to the selective amplification of a SAW and a SSLW. The Rayleigh velocity and longitudinal velocity of the waves can be extracted from the position of the peak maxima, in addition to an independent measurement of the inter-pulse delay. At resonance, the temporal full width at half maximum (FWHM) of a SSLW is smaller than the FWHM of a SAW, as evidenced from the higher amplitude of the SSLW resonance in~Fig.~\ref{fig:phase_matching}; this stems from the fact that the longitudinal wave velocity is about twice the Rayleigh wave velocity.

\section*{Phase contrast enhancement by defocusing}
	\label{defocus}
The following derivation is based on~\cite{Koehl_1999}. Consider a phase object~$\phi(x, y)$ at~$z = 0$ illuminated by a monochromatic plane wave of wavelength~$\lambda$. The field after the object is given by
\begin{equation}
	\begin{split}
		u(x, y, 0)
		= e^{i \phi(x, y)}.
	\end{split}
\end{equation}
It is assumed that~$|\phi(x, y)| \ll 1$ at every point, such that $u(x, y, 0) \approx 1 + i \phi(x, y)$. If the phase object is a single sinusoidal function, that is, $\phi(x, y) = A \cos{(k_0 x)}$, its profile can be fully converted to an amplitude profile by a slight propagation of the field~$u(x, y, z)$ due to Fresnel diffraction (this is the Talbot effect). For an arbitrary phase profile, each Fourier component gets diffracted, and the phase profile will be partially converted to an amplitude profile as the field propagates along~$z$. For this case, the field at $z = 0$ can be decomposed into its Fourier components:
\begin{equation}
	\begin{split}
		u(x, y, 0)
		= 1 + \frac{i}{(2\pi)^2} \int_{-\infty}^{\infty}\!dx \int_{-\infty}^{\infty}\!dy\, \hat{\phi}(k_x, k_y)\, e^{i \left(k_x x + k_y y\right)}.
	\end{split}
\end{equation}
For each Fourier component, the Fresnel integral gives, in the paraxial limit, the evolution of the field along its propagation axis:
\begin{equation}
	\begin{split}
		u(x, y, z)
		= \frac{1}{i \lambda z}
		  \int_{-\infty}^{\infty}{\!dx' }
		  \int_{\infty}^{\infty}{\!dy' u(x', y', 0)\, e^{i k \left[(x - x')^2 + (y - y')^2\right]/2z}}.
	\end{split}
\end{equation}
In Fourier space,
\begin{equation}
	\begin{split}
		\hat{U}(k_x, k_y, z)
		= e^{-i\left[k_x^2 +k_y^2\right]z/2k}\, \hat{U}(k_x, k_y, 0),
	\end{split}
	\label{eq:fresnel_fourier}
\end{equation}
with $k = 2\pi/\lambda$.

If the field propagates for a short distance relative to all major Fourier components of the phase object, that is, $z \ll k / \max{(k_x)}^2$ and $z \ll k / \max{(k_y)}^2$, then Eq.~\ref{eq:fresnel_fourier} can be expanded in powers of $z$ to yield
\begin{subequations}
	\begin{align}
		\hat{U}(k_x, k_y, z)
	&	= \left[1 - \frac{i z}{2 k} \left(k_x^2 +k_y^2\right)\right]\, \hat{U}(k_x, k_y, 0) + \mathcal{O}(z^2)\\
	&	= \hat{U}(k_x, k_y, 0) - \frac{i z}{2 k} \left(k_x^2 +k_y^2\right) \left[\delta{(k_x)}\delta{(k_y)} + i\hat{\phi}(k_x, k_y, 0)\right].
	\end{align}
\end{subequations}
Therefore,
\begin{equation}
	\begin{split}
		u(x, y, z)
	&	= u(x, y, 0) - \frac{z}{2 k} \nabla^2{\phi(x, y, 0)}
	\end{split}
\end{equation}
The intensity of the propagating field at a distance $z$ from the object is given by
\begin{subequations}
	\begin{align}
		|u(x, y, z)|^2
	&	= 1 - \frac{z}{k} \cos{(\phi)}\, \nabla^2{\phi(x, y, 0)} + \mathcal{O}(z^2)\\
	&	\approx 1 - \frac{z \lambda}{2\pi} \nabla^2{\phi(x, y, 0)}.
	\end{align}
\end{subequations}
The contrast is given by
\begin{subequations}
	\begin{align}
		\frac{\Delta{I}}{I}
	&	= \frac{|u(x, y, z)|^2 - |u(x, y, 0)|^2}{|u(x, y, 0)|^2}\\
	&	= - \frac{z \lambda}{2\pi} \nabla^2{\phi(x, y, 0)}.
	\end{align}
\end{subequations}
It is proportional to the Laplacian of the profile of the phase object and scales linearly with the propagation distance.

\section*{Numerical simulations}
	\label{simulations}
As a short SAW pulse propagates across a nonlinear elastic medium, its shape changes gradually due to the nonlinear distortion. Within the quadratic approximation and assuming the nonlinear distortion is slow (in the sense that the change on a distance comparable with the pulse width is small), the nonlinear evolution can be understood as an interaction among the Fourier components that results in generation of sum and differential frequencies. Quantitatively, this process is described by the evolution equation system derived in~\cite{Eckl_2004}:
\begin{equation}
	\begin{split}
		i \diff{}{\tau} B_n
		= n v_R \left[
			\sum_{m = 0}^{n} F(m/n) B_{m}^{\phantom{*}} B_{n-m}^{\phantom{*}} +
			2 \sum_{m = n}^{\infty} (n/m) F^*(n/m) B_{m}^{\phantom{*}} B_{m-n}^*
		\right].
	\end{split}
\end{equation}
Here, $B_n$ is the $n$-th Fourier component of the strain of the propagating wave, $\tau$ is the propagation distance, and the function~$F$ characterizes the rate of harmonics generation.

A waveform recorded at a location close to the source is expanded in a Fourier series; the resulting complex-valued coefficients are fed into the system of coupled differential equations as initial conditions. The system is integrated numerically over the propagation distance, and finally the inverse Fourier transform gives the predicted waveform.

The $n$-th evolution equation shows variation of the $n$-th Fourier component with the propagation distance due to the three-wave interactions between harmonics $m$ and $k$ such that $m \pm k = n$. The rate of interactions between harmonics $m$ and $k$ depends on the spatial distribution of each wave, on the third-order elastic constants of the crystal, and on the anisotropy. These effects reduce to a single dimensionless kernel~$F$. Note that $F(X)$ is symmetric about $X = 1/2$, and $F(1/2)$ quantifies the generation of the second harmonic for an initially monochromatic SAW.

\begin{figure}[h!]
\centering
	\includegraphics[width=0.5\columnwidth]{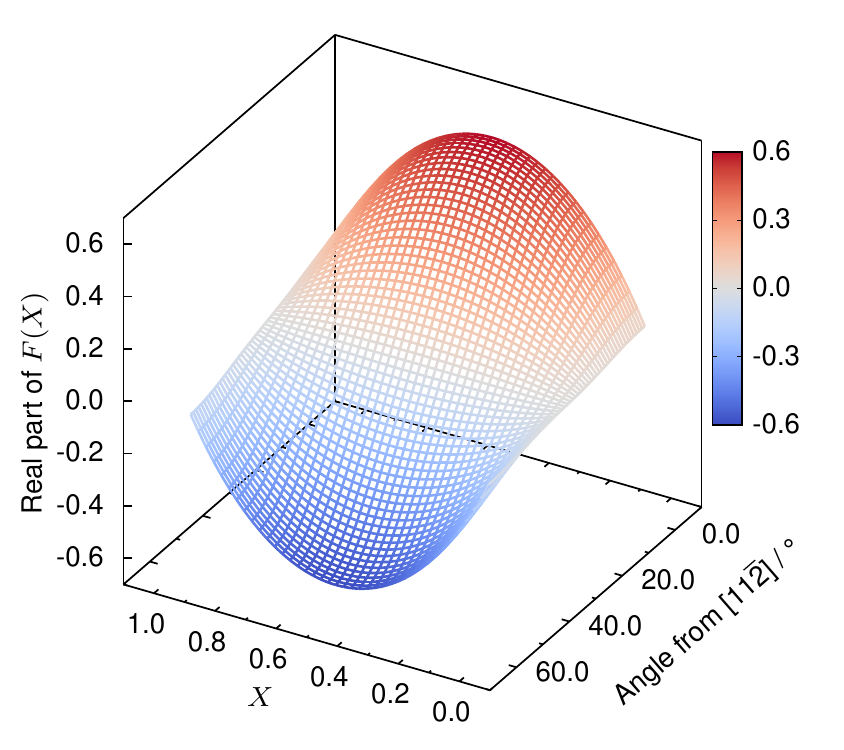}
	\caption{Real part of the kernel~$F$ for STO(111).}
	\label{fig:f_kernel}
\end{figure}

The characteristic length of the nonlinear distortions can be defined as
\begin{equation}
	\begin{split}
		x_{NL}
		= \frac{\lambda}{|B_1| |F(1/2)|},
	\end{split}
\end{equation}
where $\lambda$ denotes the typical wavelength of the initial wave, and $B_1$ is amplitude.

On the (111) plane of STO, the kernel~$F$ varies with the propagation direction, as shown on Fig.~\ref{fig:f_kernel}, where it is calculated for the~$B_n$ denoting the surface inclination in the propagation direction. This figure shows that the strongest nonlinear behavior occurs along the $\text{[11}\overline{\text{2}}\text{]}$ direction and its reverse, $\text{[}\overline{\text{11}}\text{2]}$. Note that in those directions the real parts of $F(1/2)$ have opposite signs, so the character of nonlinear evolution differs significantly.

\begin{table}[h!]
\centering
\caption{Parameters used for the numerical simulations in Si and STO. The second- and third-order elastic constants are taken from \cite{Lardner_1986, Yang_2013}.}
\begin{tabular}{l l l}
\toprule
        & Si & SrTiO$_3$ \\
        \toprule
       $\rho$ / kg $\cdot$ m$^{-3}$& 2328 & 5175 \\
        $C_{11}$ / GPa & 165.64 & 294 \\
        $C_{12}$ / GPa & 63.94 & 102 \\
        $C_{44}$ / GPa & 79.51 & 116 \\
        $C_{111}$ / GPa & -795 & -4960 \\
        $C_{112}$ / GPa & -445 & -770 \\
        $C_{123}$ / GPa & -75 & 20 \\
        $C_{144}$ / GPa & 15 & -810 \\
        $C_{166}$ / GPa & -310 & -300 \\
        $C_{456}$ / GPa & -86 & 90 \\
       $v_R$ along $\text{[}\overline{\text{11}}\text{2]}$ / m $\cdot$ s$^{-1}$ & 4735 & 4129 \\
        $|F(1/2)|$ & 0.328 & 0.6276 \\
        \toprule
\end{tabular}
\end{table}

\bibliography{supplement}